\documentclass[amsmath,amssymb,12pt,showpacs]{revtex4}
\usepackage{doublespace,graphicx}

\begin{document}

\title{Continuous-variable quantum information processing with squeezed states of light}

\author{Hidehiro Yonezawa$^{1,2}$ and Akira Furusawa$^{1,2}$}
\affiliation{$^{1}$Department of Applied Physics, School of Engineering,
The University of Tokyo,\\
7-3-1 Hongo, Bunkyo-ku, Tokyo 113-8656, Japan \\
$^{2}$CREST, Japan Science and Technology (JST) Agency,
1-9-9 Yaesu, Chuo-ku, Tokyo 103-0028, Japan }

\begin{abstract} 
We investigate experiments of continuous-variable quantum information processing based on the teleportation scheme. Quantum teleportation, which is realized by a two-mode squeezed vacuum state and measurement-and-feedforward, is considered as an elementary quantum circuit as well as quantum communication. By modifying ancilla states or measurement-and-feedforwards, we can realize various quantum circuits which suffice for universal quantum computation. In order to realize the teleportation-based computation we improve the level of squeezing, and fidelity of teleportation. With a high-fidelity teleporter we demonstrate some advanced teleportation experiments, i.e., teleportation of a squeezed state and sequential teleportation of a coherent state. Moreover, as an example of the teleportation-based computation, we build a QND interaction gate which is a continuous-variable analog of a CNOT gate. A QND interaction gate is constructed only with ancillary squeezed vacuum states and measurement-and-feedforwards. We also create continuous-variable four mode cluster type entanglement for further application, namely, one-way quantum computation.
\end{abstract}
\maketitle

\section{Introduction}

Continuous-variable (CV) quantum information processing \cite{Braunstein_QICV(2003),Braunstein2005} has been considered as a promising alternative to discrete variable quantum information processing.
Particularly quantum optics offers well-established tools for CV quantum information processing, that is, linear optics, squeezed states and homodyne detections.
While discrete variable (qubit) quantum information processing depends on a single photon detector which has a limited efficiency, in CV quantum information we can achieve homodyne detections with almost unit-efficiency. Furthermore there recently have been great steps to improve the squeezing level, up to -9dB at 860nm \cite{Takeno_S2007} and -10dB at 1064nm \cite{Vahlbruch_10dB2008} which show great potentials of CV quantum information processing with squeezed states. 

In quantum optical setting, an electromagnetic field mode is represented by an annihilation operator $\hat{a}$ with real and imaginary parts $\hat{x}$ and $\hat{p}$ corresponding to the position and momentum quadrature-phase amplitude operators. These operators $\hat{x}$ and $\hat{p}$ satisfy the commutation relation $[\hat{x},\hat{p}] = i \hbar $ (we set $\hbar=1$).
CV quantum information processing is achieved by a unitary operation $\hat U=e^{-i\hat H t/ \hbar}$ upon these variables $\hat x \rightarrow \hat U^\dagger \hat x \hat U$ ($\hat p \rightarrow \hat U^\dagger \hat p \hat U$). 
Note that a unitary operation $\hat U$ is also called a \textit {quantum gate} in the context of quantum information processing, and here $\hat H$ is the corresponding Hamiltonian. Universal quantum computation is realized by an arbitrary unitary operation. Hence one of our goals will be the implementation of an arbitrary unitary operation. Lloyd and Braunstein demonstrated that an arbitrary Hamiltonian is constructed with arbitrary quadratic Hamiltonians and another Hamiltonian which has third or higher order \cite{Lloyd_UCVQC1999}. 
Their proposal is based on a simple formula $e^{i \hat H_A\delta t }e^{i \hat H_B\delta t }e^{-i \hat H_A\delta t }e^{-i \hat H_B\delta t} =e^{i \left[ \hat H_A, \hat H_B \right] \delta t^2 }+O(\delta t^3)$ (If we have two Hamiltonians $\hat H_A$ and $\hat H_B$, we can construct Hamiltonian $\left[ \hat H_A, \hat H_B \right]$.). 
Here quadratic Hamiltonians correspond to Gaussian operations, namely, 
displacement $\hat H_d =\hat x \ (or \ \hat p)$, phase shift $\hat H_{ps} =\left( \hat x^2+ \hat p^2 \right)/2$, squeezing $\hat H_{sq} =\left( \hat x\hat p +\hat p\hat x \right)/2$, and beam splitter $\hat H_{BS} =\left( \hat p_1 \hat x_2 - \hat x_1 \hat p_2 \right)/2$. These Gaussian operations can be efficiently performed with linear optics and squeezers. 
Therefore we need to realize arbitrary Gaussian operations and one of non-Gaussian operations for universal quantum computation.
  
In order to construct fault-tolerant quantum gates we can exploit quantum teleportation \cite{Bennett93,Vaidman94,Braunstein98}.
Although quantum teleportation was first proposed in the context of quantum communication \cite{Bennett93}, it is also used as a building block in quantum computation.
Quantum teleportation is characterized by an ancilla state preparation and measurement-and-feedforward. 
It is known that by modifying the entangled resource of teleportation, some quantum gates could be implemented (off-line scheme) \cite{Gottesman99,Zhou_Methodology00,Bartlett_QTofOQG03}. Moreover, by choosing appropriate measurement-and-feedforwards with a fixed entangled resource (cluster state) we can realize universal one-way quantum computation \cite{Raussendorf_OWQC01,Menicucci06,Menicucci08}.

The teleportation-based computations are illustrated in fig. 1.
Figure 1 (a) shows a generalized teleportation circuit \cite{Zhou_Methodology00,Loock_ExampleGCC07}. A Quantum Non-Demolition (QND) gate, which is a CV analog of a CNOT gate, is realized with a Hamiltonian $\hat H=\hat x_{in} \hat p_a$. 
Off-line scheme (ancilla state preparation) is shown in fig. 1 (b). In this case, first we apply a gate $\hat U$ to the teleportation output, and then move the gate $\hat U$ before the QND gate.
Note that, in general, $\hat U$ doesn't commute with a displacement gate and a QND gate. But some important gate, e.g., a cubic phase gate $\hat U_{cp}=\exp (i \gamma \hat p_a ^3)$ \cite{Gottesman01} commutes with a QND gate $\hat U_{QND}=\exp(i \hat x_{in} \hat p_a)$. Moreover the displacement $\hat D_p$ can be replaced by $\hat U_{cp} \hat D_p \hat U_{cp}^\dagger$ which consists of squeezing and displacement \cite{Bartlett_QTofOQG03}. Hence a cubic phase gate, which will be one of universal gate sets, can be constructed with Gaussian operations and a cubic phase state which can be fault tolerantly prepared. Furthermore off-line scheme is useful not only for non-Gaussian gate constructions but also for Gaussian gate constructions. 
For example, we can construct a squeezing gate by using off-line scheme \cite{Filip05,Yoshikawa07}. Although this scheme is not exactly the same as the teleportation-based scheme described above, it is essentially same. In order to construct the squeezing gate, we use a transmittance-variable beam splitter and an ancillary squeezed vacuum state instead of a QND gate and the ancilla state $e^{-i\hat H_{sq} t}\left | x=0 \right \rangle$ in fig.1 (b). In this scheme, we can avoid  to apply direct nonlinear process which is often lossy and experimentally difficult to control. Moreover we can control the squeezing parameter by varying transmittance of the beam splitter. This controllability of squeezing leads to construction of some other gates which consist of squeezing gates and beam splitters, e.g., a QND gate \cite{Filip05,Yoshikawa08}. Although a QND gate is often difficult to realize experimentally, realization of a QND gate would be desirable for the potential applications. 

Another teleportation-based computation (one-way computation) in fig. 1 (c) is also easily understood. In this case, first we apply a gate $\hat U$ to the input, and then move the gate in front of the measurement. Again we consider some unitary gate which commutes with a QND gate. Then the unitary gate can be replaced by a generalized measurement $\hat U^\dagger \hat p \hat U$. Hence by applying a sequence of generalized teleportation with the generalized measurements $\hat U_1, \hat U_2, \cdots, \hat U_n$, we can realized $\left| \psi \right \rangle \rightarrow \hat U_n \cdots \hat U_2 \hat U_1 \left| \psi \right \rangle$. In this scheme, we only need to change measurement bases. Because the ancilla state is fixed through operations, we can prepare all the ancilla states in advance, that is a cluster state \cite{YukawaCluster08,Menicucci07,Zaidi08}, before computation. In one-way computation, universality is fulfilled by applying non-Gaussian measurements.

In this report, we present high-fidelity teleportation experiments and some examples of the teleportation-based computation, namely, a QND gate and generation of a cluster state.

\section{High-fidelity quantum teleportation}
In CV teleportation \cite{Braunstein98,Furusawa98} we exploit entanglement which can be generated by combining two squeezed vacua on a half beam splitter. 
Because of the finite level of squeezing, entanglement has always finite correlation, 
 which leads to imperfect teleportation. In order to realize the teleportation-based CV computation, it is of great importance to improve the performance of teleportation which is often evaluated with fidelity $F=\left \langle \psi_{in} \right| \hat \rho_{out} \left | \psi \right \rangle$. Here $\left | \psi \right \rangle$ is an input state and $\hat \rho_{out}$ is a density operator for the teleported state.
Since the first realization of teleportation ($F=0.58$ \cite{Furusawa98}), fidelity has been improved \cite{Bowen03a,Zhang03,Takei05e,Yonezawa_QTS(2007),Yukawa08}.
Thanks to the recent improvement of squeezing (-9dB at laser wavelength 860nm \cite{Takeno_S2007}), we obtained fidelity 0.83 for teleportation of a coherent state \cite{Yukawa08}. Fidelity 0.83 corresponds to capability of five sequential teleportations, i.e., after five times of teleportation the fidelity will be still beyond the classical limit 1/2 \cite{Braunstein00,Hammerer05}. 
Repetition of teleportation is essential for the teleportation-based computation, i.e., one-way computation. In one-way computation, computation is achieved by changing measurement bases in each teleportation step.

As high-fidelity quantum teleportation experiments, we have realized sequential teleportation of a coherent state \cite{Yonezawa_SQT(2007)} and teleportation of a squeezed state \cite{Yonezawa_QTS(2007)}.
In order to implement one-way quantum computation, we need to perform teleportation repeatedly. Hence it is of practical importance to demonstrate sequential teleportation experimentally. Moreover teleportation should be applied to an arbitrary input state including non-classical states, e.g., a squeezed state. Here squeezing below vacuum noise level indicates entanglement between the corresponding frequency sidebands. Hence observing squeezing in the teleported state indicates success of teleportation of entanglement. In this section, both experimental results of sequential teleportation and teleportation of a squeezed state will be shown.

In our experiment, we use optical parametric oscillators (OPOs) which contain periodically poled KTiOPO$_4$ as a nonlinear medium \cite{Takeno_S2007}. The output of a Ti:sapphire laser at 860nm is frequency doubled in an external cavity containing a 10mm long potassium niobate crystal. The output beam at 430nm is injected to OPOs. The output squeezed states of two OPOs are used to generate the EPR beams. Here we use a frequency sideband at 1MHz.

In the following, we describe the teleportation process \cite{Braunstein98,Furusawa98,Bowen03a,Zhang03,Takei05e,Yonezawa_QTS(2007),Yukawa08} in the Heisenberg representation. Initially, the sender Alice and the receiver Bob share a pair of EPR beams. Alice performs a joint measurement on her EPR mode ($\hat x_{\rm A}$, $\hat p_{\rm A}$) and the input mode ($\hat x_{in}$, $\hat p_{in}$). She combines these two modes at a half beam splitter and measures $\hat x_u=\left( \hat x_{in} - \hat x_{\rm A} \right)/\sqrt{2}$ and $\hat p_v=\left( \hat p_{in} + \hat p_{\rm A} \right)/\sqrt{2}$ with two homodyne detectors. The measurement results ($x_u$, $p_v$) are then sent to Bob through classical channels with gain
$g_x$ and $g_p$. For simplicity, these gains are fixed throughout the experiment and treated as unity.

Bob receives Alice's measurement results ($x_u$, $p_v$) through
the classical channels and displaces his EPR beam ($\hat x_{\rm
B}$, $\hat p_{\rm B}$) accordingly, $ \hat x_{\rm B} \rightarrow
\hat x_{out}=\hat x_{\rm B} +\sqrt{2}x_u $ and $ \hat p_{\rm B}
\rightarrow \hat p_{out}=\hat p_{\rm B} +\sqrt{2}p_v $. 
The teleported mode can be written as \cite{Takei05e}
\begin{align}
\hat x _{out}=\hat x_{in} -(\hat x_{\rm A}-\hat x_{\rm B}), \ \
\hat p _{out}=\hat p_{in} +(\hat p_{\rm A}+\hat p_{\rm B}).
\end{align}
Ideally, the EPR beams would have perfect correlations such that 
$\hat x_{\rm A}-\hat x_{\rm B} \rightarrow 0$ and $\hat p_{\rm A}+\hat p_{\rm B} \rightarrow 0$. Hence, the teleported output  would be identical to the input.
In a real experimental situation, EPR beams have finite correlation and the variance would be written as $\left \langle \left [ \Delta (\hat x_{\rm A}-\hat x_{\rm B}) \right]^2 \right \rangle = \left \langle \left [ \Delta (\hat p_{\rm A}+\hat p_{\rm B}) \right]^2 \right \rangle=2e^{-2r}\sigma_{vac}$. Here $\sigma_{vac}$ is a variance of vacuum fluctuation, and r is a squeezing parameter. 
Hence these additional noise terms are inevitably added to the output. The performance of CV quantum teleportation is determined by these noise terms. 
In order to improve fidelity of teleportation, we improve the level of squeezing $r$.
In addition, we improve mechanical stability of the setup. This is because the imperfect phase locking makes effective squeezing level lower \cite{Takeno_S2007}.

In the experiment of teleportation of a squeezed state \cite{Yonezawa_QTS(2007)}, we use three OPOs and hence three squeezed vacua. One of three squeezed vacua is used as an input. 
Fig 2 (a) shows the input squeezed state. The variances of the input state's  squeezed ($x$) and anti-squeezed quadratures ($p$) are $-6.2 \pm 0.2$dB and $12.0 \pm0.2$dB, respectively. The result of teleportation is shown in fig. 2 (b).  The measured variances of the output state are $-0.8 \pm 0.2$dB and $12.4 \pm 0.2$dB for the $x$ and $p$ quadratures, respectively. These levels of sub-vacuum noise measured in the output clearly demonstrate that the squeezing (and hence the entanglement) is preserved in the process of teleportation.

In the experiment of sequential teleportation of a coherent state \cite{Yonezawa_SQT(2007)}, we use four OPOs and construct two teleportation setup. The experimentally obtained Wigner functions \cite{Leohnardt97,Breitenbach97} are shown in fig. 3. 
The input coherent state was teleported twice and the individual teleportation fidelities are evaluated as $F_{1}=0.70\pm0.02$ and $F_{2}=0.75\pm0.02$, while the fidelity between the input and the sequentially teleported states is determined as $F^{(2)}=0.57\pm0.02$. This still exceeds the optimal classical teleportation fidelity $F_{cl}=1/2$  and almost attains the value of the first (unsequential) quantum teleportation experiment with coherent states.

\section{A quantum non-demolition gate}
A Quantum non-demolition (QND) gate plays an important role in CV quantum information processing. This is because a QND gate with unity interaction gain is a CV analog of a CNOT gate \cite{Yoshikawa08,Bartlet02}. That is to say, we can make interaction between two optical modes with a QND gate in the same manner as the qubit case. Therefore a QND gate is considered as one of the most fundamental CV quantum gates. 
A QND gate makes interaction between two input modes with a QND interaction Hamiltonian $\hat H_{QND}=\hat x_1 \hat p_2$ \cite{Yoshikawa08}. 
We can calculate output modes with a unitary operator $\hat U_{QND}=e^{i\hat x_1\hat p_2}$ (e.g., $\hat U_{QND}^\dagger \hat x_1^{in} \hat U_{QND}=\hat x_1^{out}$). 
Input and output relation is obtained as, 
\begin{eqnarray}
\hat x_1^{out}&=&\hat x_1^{in} \nonumber \\
 \hat x_2^{out}&=&\hat x_2^{in}+G\hat x_1^{in} \nonumber \\
\hat p_1^{out}&=&\hat p_1^{in}-G\hat p_2^{in} \nonumber \\
  \hat p_2^{out}&=&\hat p_2^{in}.
\label{eq:QNDinout}
\end{eqnarray} 
Here $G$ is an interaction gain.
Eq. (\ref{eq:QNDinout}) implies the reason why this interaction is called as \textit{quantum non-demolition}. When we focus on $x$ quadrature, $\hat x_1$ is preserved through interaction, on the other hand $\hat x_2$ has the information of $\hat x_1$. We can get the information of $\hat x_1$ without demolishing the $\hat x_1$. In this case $\hat x_2$ works as a meter for a signal $\hat x_1$.
Notably the important aspect is that QND interaction has duality for $x$ and $p$ quadratures.
In the case of $p$ quadrature, $\hat p_1$ and $\hat p_2$ are a meter and a signal, respectively. 

Although a QND interaction gate is naturally used for QND measurements, QND interaction is not necessarily needed to realize QND measurement. 
Conventional QND measurement experiments consider only one quadrature $x$ or $p$.
Here let us mention about the interaction gain $G$.
In the context of QND measurement, the bigger interaction gain is better. 
Unity interaction gain $G=1$, however, is more important for CV quantum information processing because of the correspondence to the CNOT gate.
Hereafter we set the interaction gain $G$ as unity $G=1$.

Notably it's worth noting the difference between a QND gate and a beam splitter which is the most basic tool in quantum optics. 
Beam splitters are often used to make interaction between two optical modes instead of a true CV CNOT gate. 
Beam splitter, however, cannot preserve signal quadrature through interaction. Then we cannot achieve QND measurement with a beam splitter. Moreover the most different feature is the entangling property.
It is well known that if we inject a non-classical state, e.g., a squeezed state, into a beam splitter, the output beams are in an entangled state. Beam splitter itself, however, does not have entangling property, hence cannot entangle classical states like a coherent state. Surprisingly a QND interaction gate enables us to entangle even coherent states. This feature is quite contrast to a beam splitter. This is because a QND gate itself has the entangling property. 
Because of these superior properties of a QND gate, a QND gate would be of great importance for potential applications in CV quantum information processing.

In our experiment, we use coherent states as inputs.
In order to verify the behavior of the QND gate, we check input-output relations and output-output correlations. 
A QND interaction gate should work as QND measurement for both $x$ and $p$ quadratures. Hence we check QND criteria for both $x$ and $p$ quadratures. QND criteria are defined with transfer coefficients and conditional variances as \cite{Holland90},
\begin{eqnarray}
  T_s +T_m &>& 1 \label{eq:TSTM} \\
  V\left( x_s^{out} | x_m^{out} \right) &<& 1 \label{eq:conditionalV}
\end{eqnarray} 
Here $T_s $ is the transfer coefficient for the input signal to the output signal $T_s =C^2 \left( x_s^{in}, x_s^{out} \right)$, and $T_m$ is the transfer coefficient for the input signal to the output meter $T_m=C^2 \left( x_s^{in}, x_m^{out} \right)$. Note that $C(x_1,x_2)$ is a correlation function between $x_1$ and $x_2$, and $V(x_1 | x_2)$ is a conditional variance for $x_1$ and $x_2$. These criteria are also applied to $p$ quadrature. In addition to these QND criteria, we also verify entanglement between outputs, and check the entangling property of a QND gate.

In the experiment, we realize a QND interaction gate by using off-line scheme \cite{Filip05,Yoshikawa08}.
The schematic of the experimental setup is shown in fig. 4 (a). Here we use two off-line squeezers with squeezing parameter of which can be controlled by beam splitting ratio with suitable feedback gain. The outputs can be written with the reflectance of beam splitters R as \cite{Filip05,Yoshikawa08}, 
\begin{eqnarray}
 \hat x_1^{{\rm{out}}}  &=& \hat x_1^{{\rm{in}}}  - \sqrt {\frac{{1 - R}}{{1 + R}}} \hat x_{\rm{A}}^{{\rm{(0)}}} e^{ - r_{\rm{A}} }  \nonumber \\ 
 \hat x_2^{{\rm{out}}}  &=& \hat x_2^{{\rm{in}}}  + \left( {\frac{1}{{\sqrt R }} - \sqrt R } \right)\hat x_1^{{\rm{in}}}  + \sqrt {R\frac{{1 - R}}{{1 + R}}} \hat x_{\rm{A}}^{{\rm{(0)}}} e^{ - r_{\rm{A}} }  \nonumber \\ 
 \hat p_1^{{\rm{out}}}  &=& \hat p_1^{{\rm{in}}}  - \left( {\frac{1}{{\sqrt R }} - \sqrt R } \right)\hat p_2^{{\rm{in}}}  + \sqrt {R\frac{{1 - R}}{{1 + R}}} \hat p_{\rm{B}}^{{\rm{(0)}}} e^{ - r_{\rm{B}} }  \nonumber \\ 
 \hat p_2^{{\rm{out}}}  &=& \hat p_2^{{\rm{in}}}  + \sqrt {\frac{{1 - R}}{{1 + R}}} \hat p_{\rm{B}}^{{\rm{(0)}}} e^{ - r_{\rm{B}} } 
 \end{eqnarray}

Here $r_A$ and  $r_B$ are squeezing parameters for ancilla squeezed vacuum states A and B. For unity gain we set $ {\frac{1}{{\sqrt R }} - \sqrt R }=1$, namely, R=0.38. Note that the outputs have additional noise term due to finite squeezing of ancilla states.

The experimental results are shown in fig. 4 (b), (c) and (d).
Figure 4 (b) shows the measurement results of $\hat x_1$ and $\hat x_2$ when we use a coherent state with an amplitude along $x$ quadrature for mode 1 and vacuum state for mode 2. Here $p$ quadratures are not shown because $p$ quadratures remain unchanged up to the additional noise term. 
Figure 4 (b) shows that the signal $x_1$ information is transferred to the meter $x_2$. Moreover the amplitudes of $\hat x_{1}^{in}$, $\hat x_{1}^{out}$ and $\hat x_{2}^{out}$ are almost same which assure the unity interaction gain ($G=1$). 
We also measure the variance of $x_1^{out}$ and $x_2^{out}$ by using two vacuum states as inputs. We calculate transfer coefficients which can be expressed as $T_s=SNR_s^{out}/SNR_s^{in}$ and  $T_m=SNR_m^{out}/SNR_s^{in}$ \cite{BachorAGuide}. Here SNR is Signal to Noise Ratio, hence  $T_s= \left \langle \left( \Delta x_1^{in} \right)^2 \right \rangle / \left \langle \left( \Delta x_1^{out} \right)^2 \right \rangle$ and $T_m= \left \langle \left( \Delta x_1^{in} \right)^2 \right \rangle/ \left \langle \left( \Delta x_2^{out} \right)^2 \right \rangle$. 
We obtain $T_s=0.79\pm0.03$ and $T_m=0.41\pm0.02$, then $T_s+T_m=1.20 \pm0.05 >1$ which satisfy QND criteria.
Figure 4 (c) shows the measurement results of $\hat p_1$ and $\hat p_2$ when we use a coherent state with an amplitude along $p$ quadrature for mode 2 and vacuum state for mode 1. Here $x$ quadratures are not shown again. In this figure, it can be seen that the signal $p_2$ information is transferred to the meter $p_1$. The calculated transfer coefficients are $T_s=0.71\pm0.03$ and $T_m=0.39\pm0.02$, then $T_s+T_m=1.10 \pm0.05 >1$ which satisfy QND criteria. Hence for both $x$ and $p$ quadrature QND criteria of eq. (\ref{eq:TSTM}) are verified.

Figure 4 (d) shows the output correlation for $x$ quadrature ($p$ quadrature not shown). Here we show the variance of $\hat x_1^{out} - k \hat x_2^{out}$ with an electric gain $k$. Conditional variance corresponds to the minimum value of the variance of $\hat x_1^{out} - k \hat x_2^{out}$. We obtain the conditional variance as $V\left( x_1^{out} | x_2^{out} \right)=0.75\pm0.01<1$ (similarly we measure and obtain $V\left( p_2^{out} | p_1^{out} \right)=0.78\pm0.01 <1 $). Therefore QND criteria of eq. (\ref{eq:conditionalV}) are satisfied for both $x$ and $p$ quadrature. 
Finally we verify the entanglement in the output. Sufficient condition of entanglement can be written as \cite{Duan00,Simon00}
\begin{equation}
\left \langle \left[ \Delta \left( \hat x_1^{out} - k \hat x_2^{out} \right ) \right]^2 \right \rangle<2k \ {\rm and} \ \left \langle \left[ \Delta \left( \hat p_2^{out} + k \hat p_1^{out} \right ) \right]^2 \right \rangle<2k. 
\end{equation}
Figure 4 (d) shows that the variance of $x_1^{out} - k \hat x_2^{out}$ is clearly below $2k$ in a certain $k$. We also verify $p_2^{out} + k \hat p_1^{out}$ is below $2k$ simultaneously (not shown). These results show entanglement in the output. 
Therefore we successfully demonstrate the QND interaction gate.

\section{Generation of cluster states}
In this section we show the experimental results of generation of CV four mode  cluster states. The details of the experiment are shown in ref \cite{YukawaCluster08}. 
Cluster states are resources of one-way quantum computation \cite{Raussendorf_OWQC01,Menicucci06,Menicucci08} which is another computation model than the conventional circuit model. In the circuit model, computation is represented by a sequence of quantum gates. In the one-way computation, measurements on entangled states will play a key role. The quantum computation is specified by the choice of the measurement bases and the property of the entangled states (cluster states). In the computation, due to the measurement, we cannot use resource states again. Hence this computation is irreversible in contrast to the circuit model. This is why this scheme is called one-way computation. 

A cluster state is a multipartite entangled state which may differ from even GHZ or W state. CV cluster states are created by using squeezed vacuum states and beam splitters. CV $N$-mode cluster states are defined as $N$-mode Gaussian states whose certain quadratures have perfect correlations in the limit of infinite squeezing,
\begin{equation}
\hat p_a  - \sum\limits_{b \in N_a } {\hat x_b }  \to 0,\quad (a = 1, \cdots N)
\end{equation}
Here $N_a$ are the neighboring modes of $a$. In the limit, the cluster state becomes a simultaneous zero eigenstate of these quadrature combinations.
In our experiment we create four mode cluster states. By the definition there are several types of four mode cluster states as illustrated in fig. 5 (a). 
In our experiment, we create 3 different types of four mode cluster states, linear, T-shape and diamond-shape cluster states. For example, linear cluster state is given as, 
\begin{eqnarray}
 \hat p_1  - \hat x_2 \quad \quad  &\to& 0 \nonumber \\ 
 \hat p_2  - \hat x_1  - \hat x_3  &\to& 0 \nonumber \\ 
 \hat p_3  - \hat x_2  - \hat x_4  &\to& 0 \nonumber \\ 
 \hat p_4  - \hat x_3 \quad \quad  &\to& 0. \label{eq:Ideal-linearCluster}
 \end{eqnarray} 

In order to generate cluster states, there will be several possible ways. Here we use four squeezed vacuum states and three beam splitters \cite{YukawaCluster08}. 
For example, the schematic of a setup for four mode linear cluster state \cite{vanLoock07} is shown in fig. 5 (b). We use four p-squeezed vacuum states and $\hat F$ indicates Fourier transform operator which corresponds to $-90$ degree rotation in the phase space. 
The output four modes are written with a squeezing parameter $r$,
\begin{eqnarray}
 \hat p_1  - \hat x_2 \quad \quad  &=& \sqrt 2 \hat p_{\rm A}^{(0)} e^{ - r}  \nonumber \\ 
 \hat p_2  - \hat x_1  - \hat x_3  &=& \left( {\sqrt {\frac{5}{2}} \hat p_{\rm C}^{(0)}  + \frac{1}{{\sqrt 2 }}\hat p_{\rm D}^{(0)} } \right)e^{ - r}  \nonumber \\ 
 \hat p_3  - \hat x_2  - \hat x_4  &=& \left( {\frac{1}{{\sqrt 2 }}\hat p_{\rm A}^{(0)}  - \sqrt {\frac{5}{2}} \hat p_{\rm B}^{(0)} } \right)e^{ - r}  \nonumber \\ 
 \hat p_4  - \hat x_3 \quad \quad  &=& \sqrt 2 \hat p_{\rm D}^{(0)} e^{ - r}  \label{eq:Real-linearCluster}
\end{eqnarray}
For simplicity all the squeezing parameters are set as $r$. Superscripts (0) denote the initial vacuum modes. In the limit of infinite squeezing, eq. (\ref{eq:Real-linearCluster}) become identical to the definition of the cluster state as eq. (\ref{eq:Ideal-linearCluster}). 
Furthermore this generation scheme is applicable to generation of the other cluster states. 
By changing beam splitting ratio and phase shift, we can use almost same experimental setup to create T-shape and diamond-shape cluster states.

The generated cluster state is verified by a sufficient condition for fully inseparable four mode state \cite{vanLoock03},
\begin{eqnarray}
 \left\langle \left[ \Delta \left( {\hat p_1  - \hat x_2 } \right) \right]^2  \right\rangle  + \left\langle \left[ \Delta \left( {\hat p_2  - \hat x_1  - \hat x_3 } \right) \right]^2  \right\rangle  < 4 \nonumber \\ 
 \left\langle \left[ \Delta \left( {\hat p_4  - \hat x_3 } \right) \right]^2  \right\rangle  + \left\langle \left[ \Delta \left( {\hat p_3  - \hat x_2  - \hat x_4 } \right) \right]^2  \right\rangle  < 4 \nonumber\\ 
 \left\langle \left[ \Delta \left( {\hat p_2  - \hat x_1  - \hat x_3 } \right) \right]^2  \right\rangle  + \left\langle \left[ \Delta \left( {\hat p_3  - \hat x_2  - \hat x_4 } \right) \right]^2  \right\rangle  < 4. \label{eq:inseparability} 
 \end{eqnarray} 
Here we set the variance of the vacuum mode as unity, that is, $\left \langle \left( \hat x^{(0)}  \right)^2\right \rangle =\left \langle \left( \hat p^{(0)}  \right)^2\right \rangle=1$.

The measurement result of four mode linear cluster state is shown in fig. 5 (c).
Here we show the measurement result only for $\hat p_1 - \hat x_2$. 
The variance of $\hat p_1 - \hat x_2$ is $-5.4\pm0.2$dB below the corresponding vacuum noise level, which show clear quantum correlation. We measure other variances in eq. (\ref{eq:Real-linearCluster}), and the measured correlations are more than 5dB each. 
The left terms of the inequality (\ref{eq:inseparability}) are obtained as,
\begin{eqnarray}
 \left\langle \left[ \Delta \left( {\hat p_1  - \hat x_2 } \right) \right]^2  \right\rangle  + \left\langle \left[ \Delta \left( {\hat p_2  - \hat x_1  - \hat x_3 } \right) \right]^2  \right\rangle  &=& 1.37 \pm 0.06 < 4 \nonumber \\ 
 \left\langle \left[ \Delta \left( {\hat p_4  - \hat x_3 } \right) \right]^2  \right\rangle  + \left\langle \left[ \Delta \left( {\hat p_3  - \hat x_2  - \hat x_4 } \right) \right]^2  \right\rangle &=&1.67 \pm 0.08 < 4 \nonumber\\ 
 \left\langle \left[ \Delta \left( {\hat p_2  - \hat x_1  - \hat x_3 } \right) \right]^2  \right\rangle  + \left\langle \left[ \Delta \left( {\hat p_3  - \hat x_2  - \hat x_4 } \right) \right]^2  \right\rangle &=& 1.42\pm0.07 < 4.  
 \end{eqnarray} 
Hence generation of four mode linear cluster state is verified. 
Similarly we create T-shape and diamond-shape cluster state.
Particularly diamond-shape cluster state is generated by applying local Fourier transformation to the linear cluster state. Furthermore, if we change 80\% T beam splitter with HBS in fig. 5 (b) and apply some local Fourier transformations, we can also create T-shape cluster state. 
We verify the generation of four mode T-shape and diamond-shape cluster state in the similar way as the case of linear cluster state.

Now we have been investigating one-way computation with these cluster states. 
If we can use photon counting as non-Gaussian measurement, we may realize some non-Gaussian operation which leads to universal one-way computation.

\section{Conclusion}
We demonstrate high-fidelity teleportation experiments. 
The fidelity of teleportation of a coherent state is obtained as $F=0.83\pm0.02$. We achieve advanced teleportation experiments, i.e., teleportation of a squeezed state and sequential teleportation of a coherent state. In the experiment of teleportation of a squeezed state, we observe $-0.8$dB squeezing below the vacuum noise level in the teleported state. We obtain the fidelity $F=0.57$ for sequentially teleported coherent state, which is still beyond the classical limit 1/2. 
As an important example of the teleportation-based computation, we demonstrate a QND interaction gate by using off-line scheme, i.e., ancillary squeezed vacuum modes and measurement-and-feedforwards.
Furthermore we create various CV four mode cluster states as resources of one-way quantum computation. 
These results show the potential and suitability of CV quantum computation in quantum optical setting.

      \begin{figure}[ht]
        \begin{center}
           \includegraphics[width=0.97\linewidth]{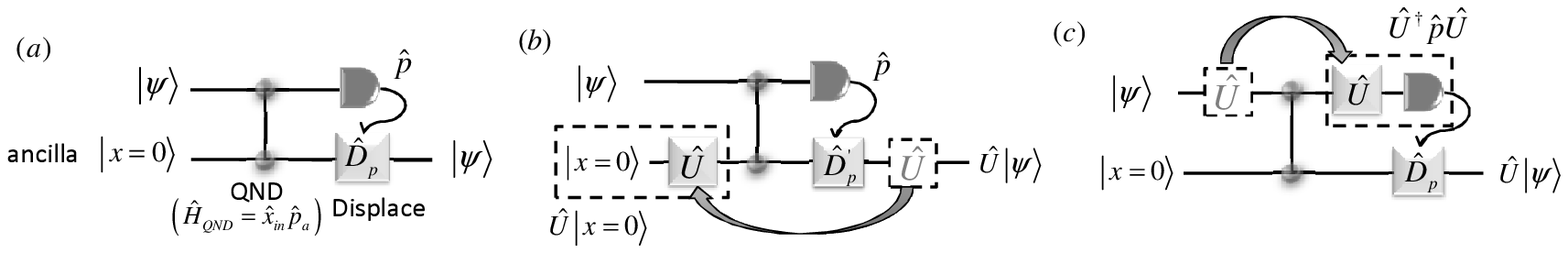}
\caption{Schematics of the generalized teleportation and the teleportation-based computation. (a) Generalized teleportation circuit. Measurement base is $\hat p$, that is, we measure $p$ quadrature in homodyne detection. $\hat D_p$ is a displacement operator for $p$ quadrature. $\left| x=0 \right \rangle$ is a position eigenstate with eigenvalue zero. A QND gate is a CV analog of a CNOT gate.
(b) Teleportation-based computation 1 (off-line scheme). Here we consider a unitary operator $\hat U$ commuting with a QND gate. $\hat D_p$ is replaced by $\hat D_p'=\hat U \hat D_p \hat U^\dagger$.
The gate operation $\hat U$ is replaced by the ancilla state preparation. (c) Teleportation-based computation 2 (one-way computation). The gate operation $\hat U$ is replaced by the generalized measurement $\hat U^\dagger \hat p \hat U$.}
           \label{fig1}
        \end{center}
\end{figure}

       \begin{figure*}[ht]
      \begin{tabular}{cc}
                \begin{minipage}{0.48\hsize}
                 \begin{flushright}
                   \includegraphics[width=0.75\linewidth,clip]{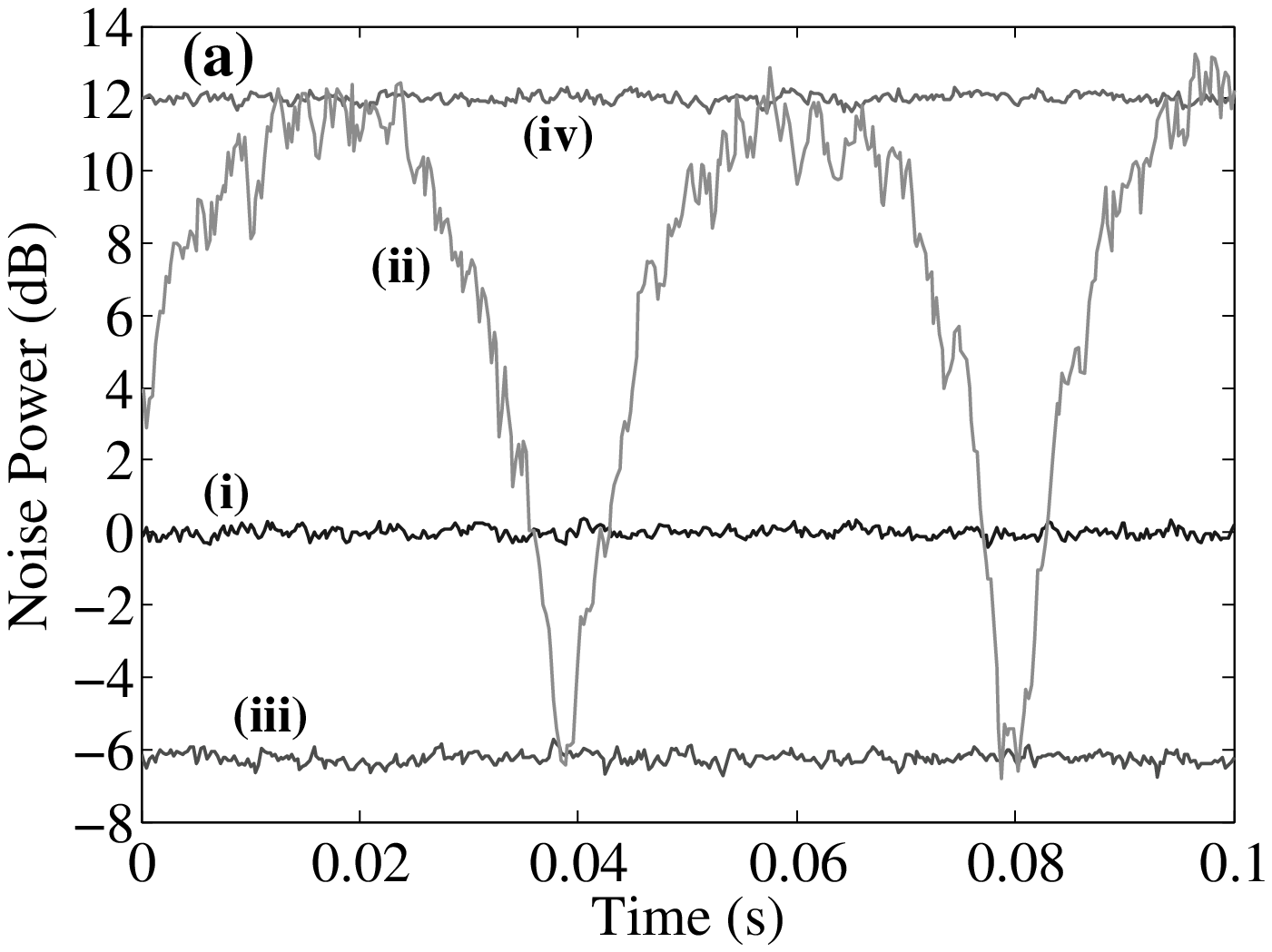}
                 \end{flushright}
                \end{minipage}
                \begin{minipage}{0.48\hsize}
                 \begin{flushleft}
                   \includegraphics[width=0.75\linewidth,clip]{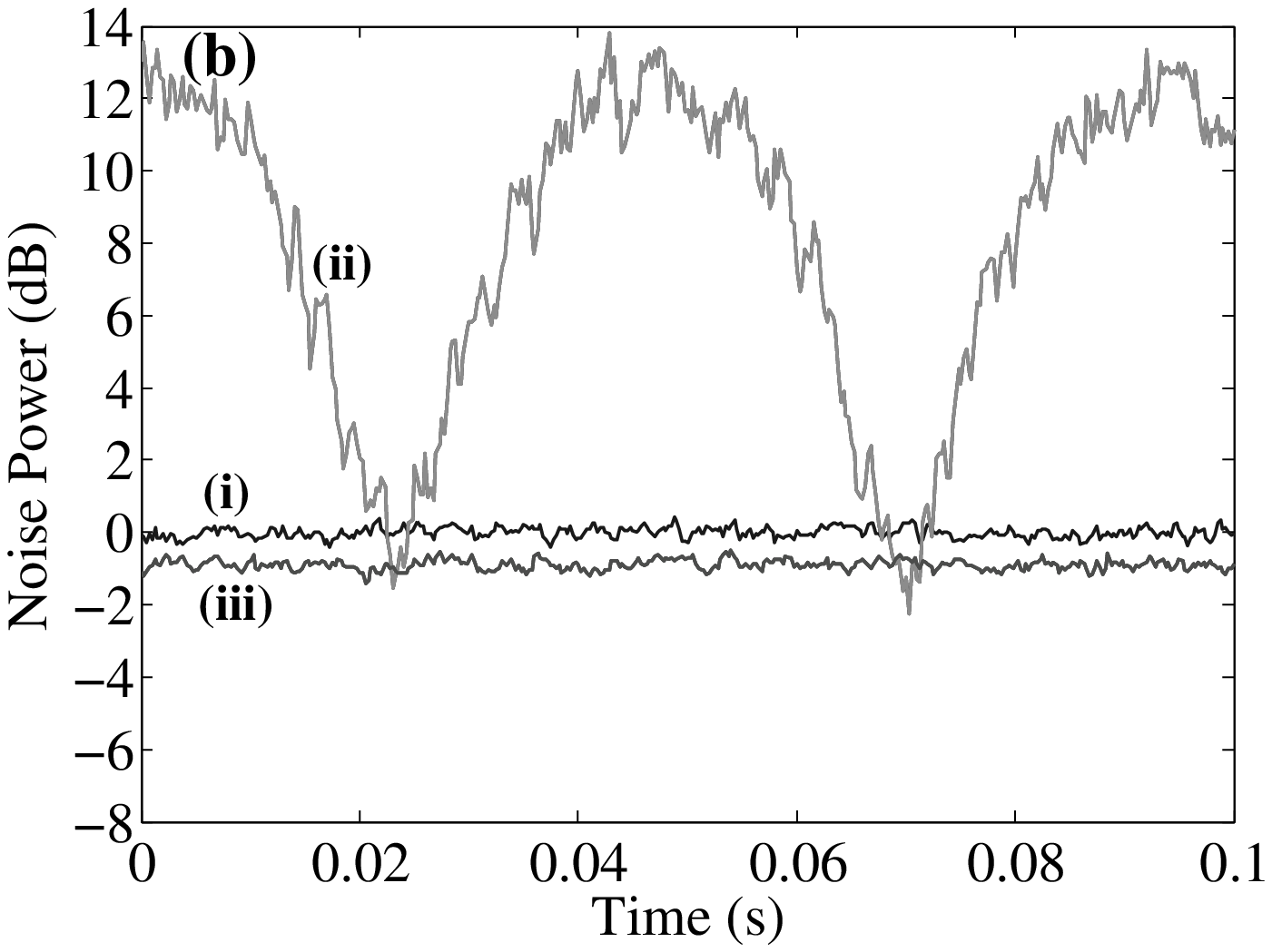}
                 \end{flushleft}
                \end{minipage} 
              \end{tabular}
\caption{
Quantum teleportation of a squeezed state. 
(a) shows the squeezed state to be teleported. (i) shows the vacuum noise level. (ii) shows the squeezed state with phase scanned. (iii) and (iv) show the squeezed state with phase locked to the squeezed and anti-squeezed quadratures. 
(b) shows the output state of the teleportation for the $x$ quadrature ($p$ quadrature not shown). (i) shows the vacuum noise level.
(ii) shows the teleported state with the input state's phase scanned. 
(iii) shows the teleported state with the input state's phase locked to the $x$ quadrature. All traces except traces (ii) are averaged 30 times. 
The center frequency is 1MHz. The resolution and video bandwidths are 30 kHz and 300 Hz, respectively.}
\label{fig2}
       \end{figure*}

       \begin{figure*}[ht]
         \begin{center}
                   \includegraphics[width=0.85\linewidth,clip]{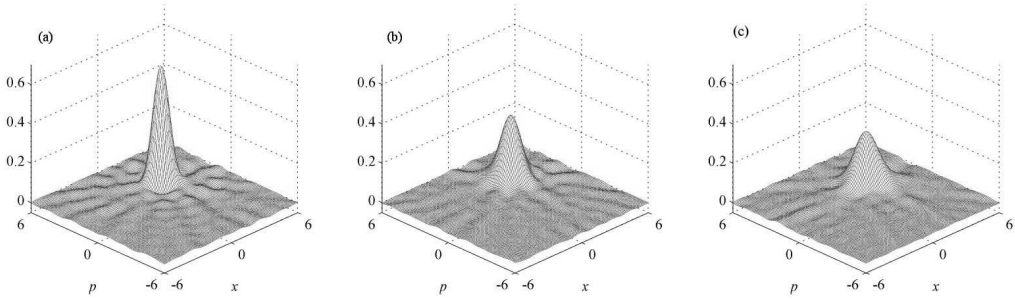}
\caption{Sequential teleportation of a coherent state. Wigner functions are reconstructed by using optical homodyne tomography. Here we use $\hbar=1/2$, then $\left \langle \left(\hat x^{(0)}\right)^2\right \rangle=\left \langle \left(\hat p^{(0)}\right)^2\right \rangle=1/4$. (a) Input coherent state. (b) Teleported state. (c) Sequentially teleported state. In these measurement, we locked the phase of input coherent state $\sim$ 45$^\circ$ from $x$ quadrature.}
\label{fig3}
       \end{center}
       \end{figure*}

       \begin{figure*}[ht]
      \begin{tabular}{ccc}
                \begin{minipage}{0.4\hsize}
                 \begin{flushright}
                   \includegraphics[width=0.95\linewidth,clip]{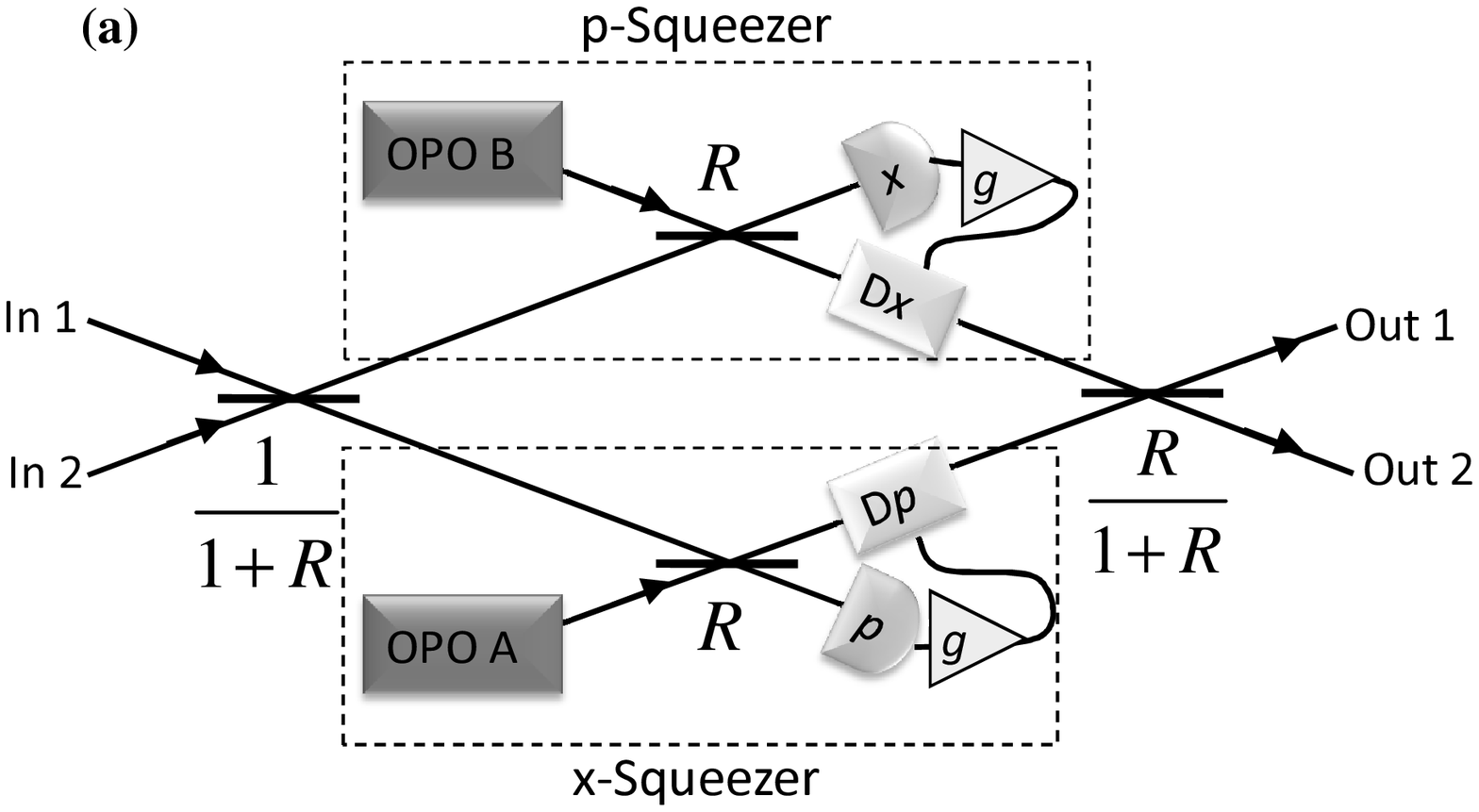}
                \end{flushright}
               \end{minipage}
                \begin{minipage}{0.6\hsize}
                 \begin{flushleft}
                   \includegraphics[width=0.9\linewidth,clip]{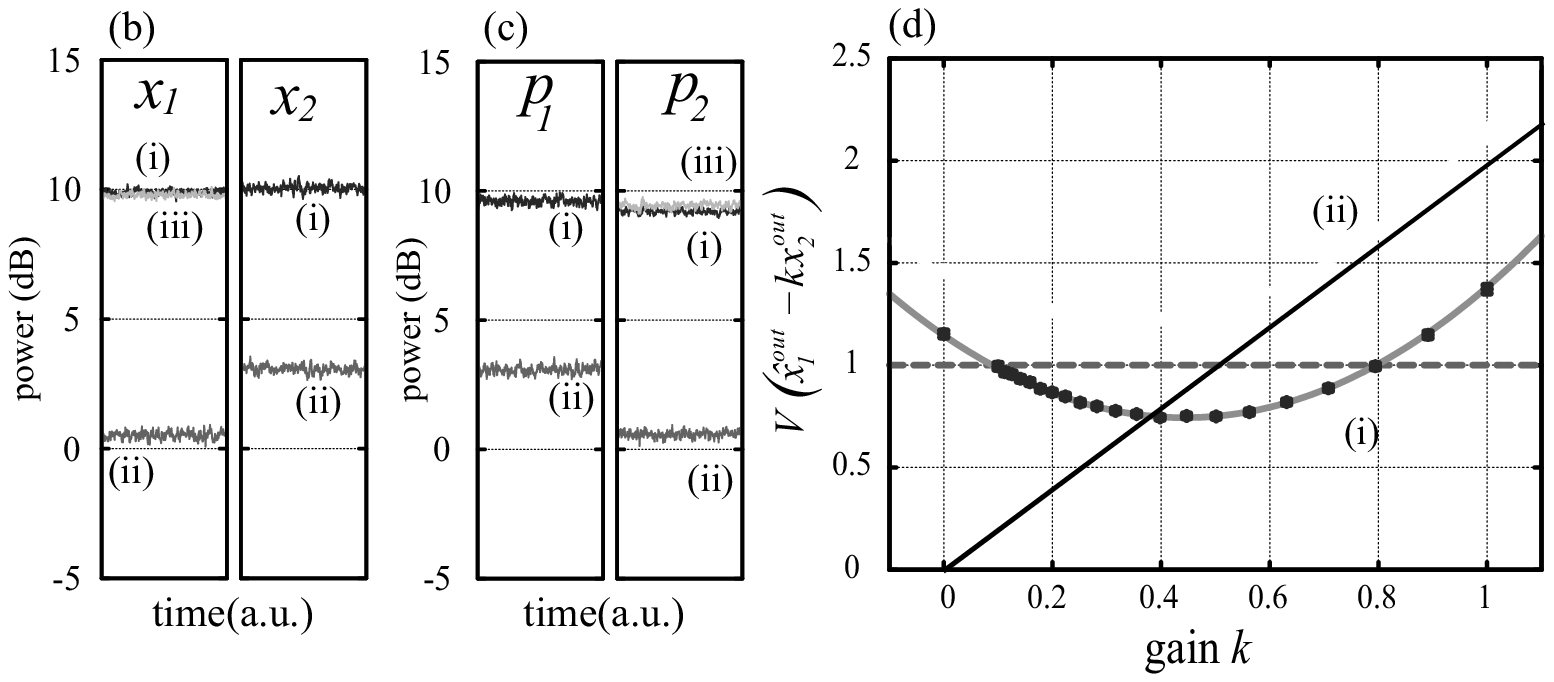}
                 \end{flushleft}
               \end{minipage} 
              \end{tabular}
\caption{A QND interaction gate. (a) Experimental set up. (b) The measurement results of $\hat x_1$ and $\hat x_2$ quadratures ($p$ quadratures not shown). Here we apply an amplitude on the signal $x_1$ quadrature. That is, $\hat x_1^{in}=\alpha_1^{in}+\hat x_1^{(0)} $, $\hat x_2^{in}=\hat x_2^{(0)} $, $\hat p_1^{in}=\hat p_1^{(0)} $ and $\hat p_2^{in}=\hat p_2^{(0)} $. Here $\alpha$ is the coherent amplitude, and superscript (0) denote vacuum modes.
The left figure in (b) shows measurement results of $\hat x_1$ quadrature. (i) the output $\left \langle \left ( \hat x_{1}^{out} \right)^2 \right \rangle$. (ii) The variance of the output $\left \langle \left ( \Delta  \hat x_{1}^{out} \right)^2 \right \rangle$. (iii) The input $\left \langle \left ( \hat x_{1}^{in} \right)^2 \right \rangle$.
The right figure in (b) shows measurement results of $\hat x_2$ quadrature. 
(i) the output $\left \langle \left ( \hat x_{2}^{out} \right)^2 \right \rangle$. (ii) The variance of the output $\left \langle \left ( \Delta  \hat x_{2}^{out} \right)^2 \right \rangle$. 
(c) The measurement results of $\hat p_1$ and $\hat p_2$ quadratures ($x$ quadratures not shown). Here we apply an amplitude on the signal $p_2$ quadrature. 
The left figure in (c) shows measurement results of $\hat p_1$ quadrature. (i) the output $\left \langle \left ( \hat p_{1}^{out} \right)^2 \right \rangle$. (ii) The variance of the output $\left \langle \left ( \Delta  \hat p_{1}^{out} \right)^2 \right \rangle$. 
The right figure in (c) shows measurement results of $\hat p_2$ quadrature. 
(i) the output $\left \langle \left ( \hat p_{2}^{out} \right)^2 \right \rangle$. (ii) The variance of the output $\left \langle \left ( \Delta  \hat p_{2}^{out} \right)^2 \right \rangle$. 
(iii) The input $\left \langle \left ( \hat p_{2}^{in} \right)^2 \right \rangle$.
(d) The variance of $\hat x_1^{out} -k \hat x_2^{out}$ as a function of the electric gain $k$. Here $V(\hat x_1^{out} -k \hat x_2^{out})$ means the variance, and $V(\hat x^{(0)})=1$
(i) Measurement results (dots) and theoretical calculation (curve). (ii) Corresponding line to 2$k$. Lowering below this line indicates the presence of the entanglement.}
\label{fig4}
       \end{figure*}

       \begin{figure*}[ht]
      \begin{tabular}{ccc}
                \begin{minipage}{0.3\hsize}
                 \begin{flushright}
                   \includegraphics[width=0.9\linewidth,clip]{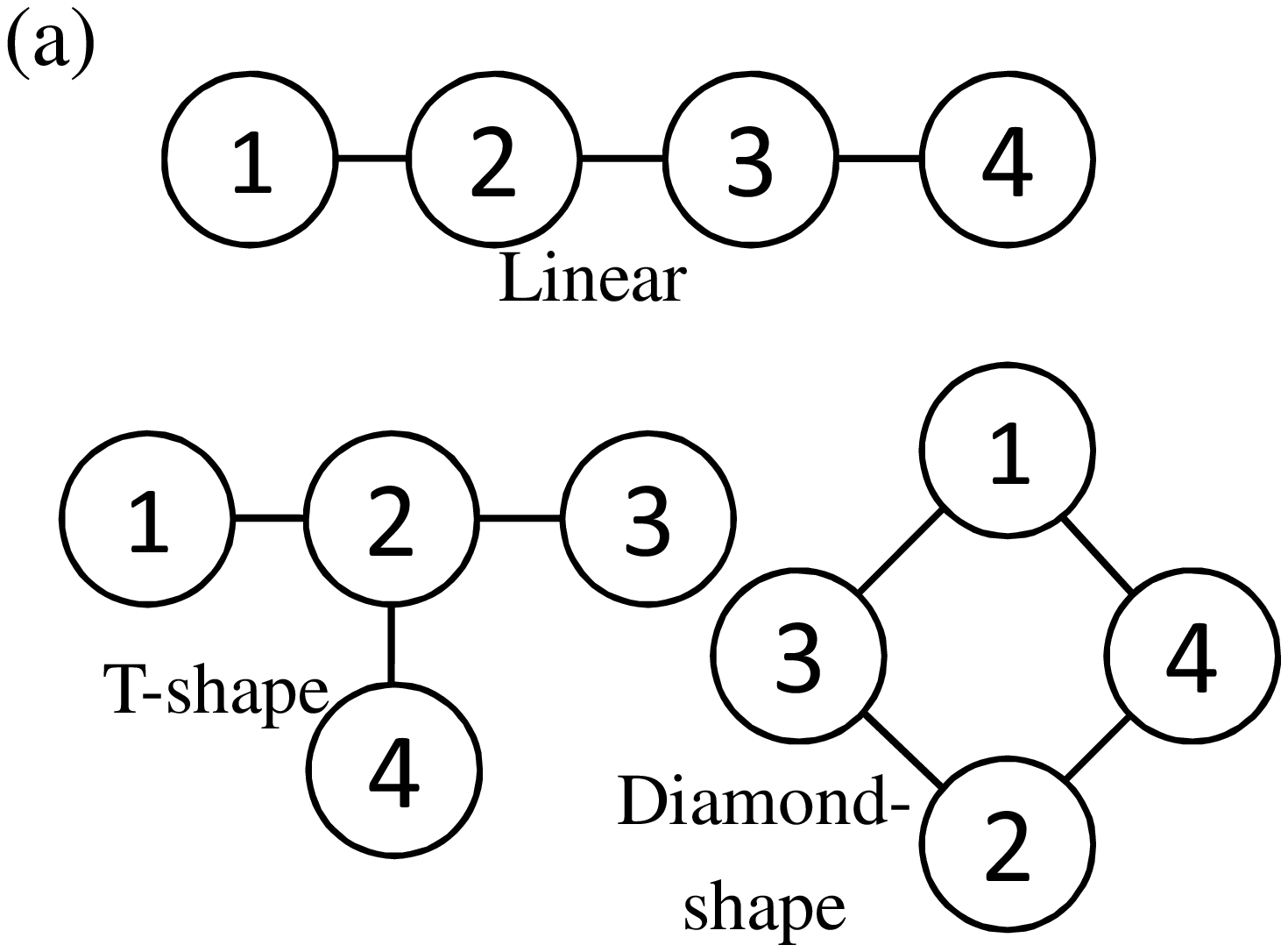}
                \end{flushright}
               \end{minipage}
                \begin{minipage}{0.3\hsize}
                 \begin{flushleft}
                   \includegraphics[width=0.9\linewidth,clip]{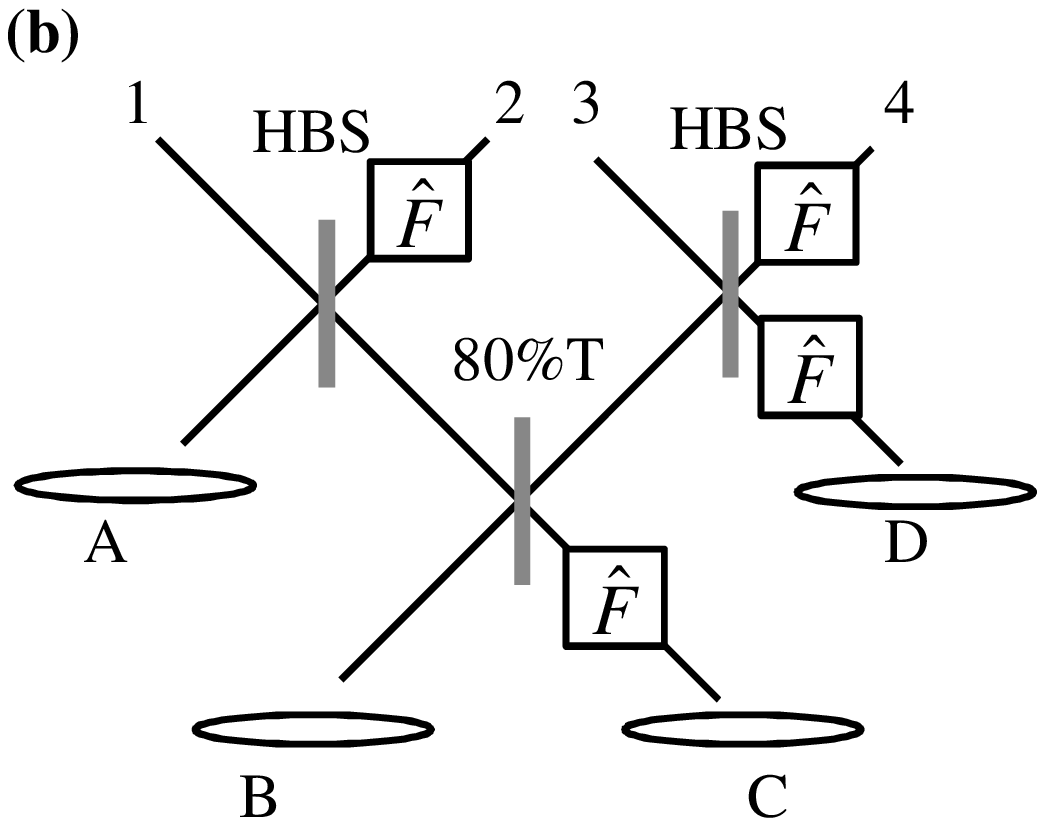}
                 \end{flushleft}
               \end{minipage} 
                \begin{minipage}{0.3\hsize}
                 \begin{flushleft}
                   \includegraphics[width=0.9\linewidth,clip]{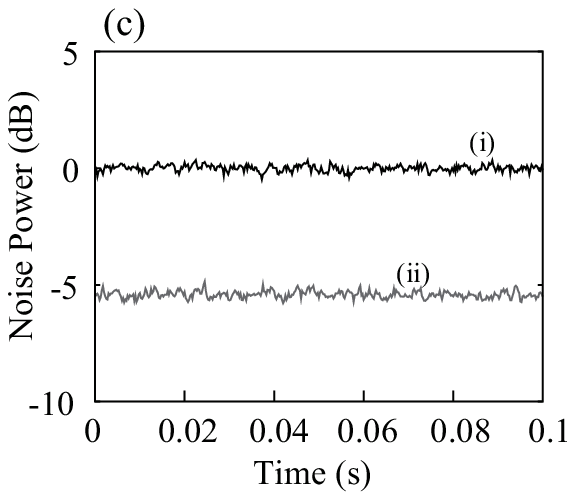}
                 \end{flushleft}
               \end{minipage} 
              \end{tabular}
\caption{Generation of cluster states. (a) Various types of four mode cluster states. Linear, T-shape, diamond-shape cluster state are shown. (b) Setup for generation of four mode linear cluster state. Ellipses represent squeezed vacua. Here all the squeezed vacua are p-squeezed state. HBS is a half beam splitter, $\hat F$ is a Fourier transform operator.
(c) Example of the measurement results. Here the variance of $\hat p_1-\hat x_2$ for linear cluster state is shown. (i) shows the corresponding vacuum noise level. (ii) shows the variance of $\hat p_1-\hat x_2$. The variance is $-5.4\pm0.02$dB below the vacuum noise level.}
\label{fig5}
       \end{figure*}


\begin{thebibliography}{}

    \bibitem{Braunstein_QICV(2003)}
      S. L. Braunstein and A. K. Pati,
      \textit{Quantum Information with Continuous Variables}
      (Kluwer Academic Publishers, Dordrecht, 2003).

    \bibitem{Braunstein2005}
      S. L. Braunstein and P. van Loock,
      Rev. Mod. Phys. {\bf 77}, 513 (2005).

    \bibitem{Takeno_S2007}
      Y. Takeno, M. Yukawa, H. Yonezawa, and A. Furusawa,
      Opt. Lett. {\bf 15}, 4321 (2007).

    \bibitem{Vahlbruch_10dB2008}
     H. Vahlbruch, M. Mehmet, S. Chelkowski, B. Hage, A. Franzen, 
     N. Lastzka, S. Gossler, K. Danzmann, and R. Schnabel,
     Phys. Rev. Lett. {\bf 100}, 033602 (2008).

    \bibitem{Lloyd_UCVQC1999}
      S. Lloyd and S. L. Braunstein,
    Phys. Rev. Lett. {\bf 82}, 1784 (1999).

    \bibitem{Bennett93}
      C.H. Bennett, G. Brassard, C. Crepeau, R. Jozsa, A. Peres, 
      and W. K. Wootters
     Phys. Rev. Lett.  {\bf 70}, 1895 (1993).

    \bibitem{Vaidman94}
      L. Vaidman,
      Phys. Rev. A. {\bf 49}, 1473 (1994).

    \bibitem{Braunstein98}
      S. L. Braunstein and H. J. Kimble,
      Phys. Rev. Lett. {\bf 80}, 869 (1998).

    \bibitem{Gottesman99}
   	D. Gottesman and I.L. Chuang,
   	Nature {\bf 402}, 390 (1999).

    \bibitem{Zhou_Methodology00}
      X. Zhou, D. W. Leung, and I. L. Chuang,
      Phys. Rev. A. {\bf 62}, 052316 (2000).

    \bibitem{Bartlett_QTofOQG03}
      S. D. Bartlett and W. J. Munro,
      Phys. Rev. Lett. {\bf 90}, 117901 (2003).

    \bibitem{Raussendorf_OWQC01}
      R. Raussendorf and H. J. Briegel,
      Phys. Rev. Lett. {\bf 86}, 5188 (2001).

    \bibitem{Menicucci06}
      N. C. Menicucci, P. van Loock, M. Gu,
      C. Weedbrook, T. C. Ralph, and M. A. Nielsen,
      Phys. Rev. Lett. {\bf 97}, 110501 (2006).

     \bibitem{Menicucci08}
      N. C. Menicucci, S. T. Flammia, and O. Pfister,
      Phys. Rev. Lett. {\bf 101}, 130501 (2008).

    \bibitem{Loock_ExampleGCC07}
      P. van Loock, 
      J. Opt. Soc. Am. B  {\bf 24}, 340 (2007).

    \bibitem{Gottesman01}
      D. Gottesman, A. Kitaev, and J. Preskill,
      Phys. Rev. A. {\bf 64}, 012310 (2001).

    \bibitem{Filip05}
      R. Filip, P. Marek, and U.L. Andersen,
      Phys. Rev. A.  {\bf 71}, 042308 (2005).

    \bibitem{Yoshikawa07}
      J. Yoshikawa, T. Hayashi,T. Akiyama, N. Takei, 
      A. Huck, U. L. Andersen, and A. Furusawa
      Phys. Rev. A. {\bf 76}, 060301R (2007).

    \bibitem{Yoshikawa08}
      J. Yoshikawa, Y. Miwa, A. Huck, U. L. Andersen, 
      P. van Loock, and A. Furusawa,
      to appear in Phys. Rev. Lett., quant-ph/0808.0551v1 

    \bibitem{YukawaCluster08}
     M. Yukawa, R. Ukai, P. van Loock, and A. Furusawa,
     Phys. Rev. A. {\bf 78}, 012301 (2008).

     \bibitem{Menicucci07}
       N. C. Menicucci, S. T. Flammia, H. Zaidi, and O. Pfister,
       Phys. Rev. A {\bf 76}, 010302 (2007).

     \bibitem{Zaidi08}
       H. Zaidi, N. C. Menicucci, S. T. Flammia, R. Bloomer, M. Pysher, 
       and O. Pfister, 
       Laser Phys. 18, 659 (2008).

    \bibitem{Furusawa98}
       A. Furusawa, J. L. S\o rensen, S. L. Braunstein, C. A. Fuchs, 
       H. J. Kimble, and E. S. Polzik,
       Science  {\bf 282}, 706 (1998).

    \bibitem{Bowen03a}
      W. P. Bowen, N. Treps, B. C. Buchler, R. Schnabel, T. C. Ralph, 
      Hans-A. Bachor, T. Symul, and P. K. Lam,
      Phys. Rev. A.  {\bf 67}, 032302 (2003).

    \bibitem{Zhang03}
      T. C. Zhang, K. W. Goh, C. W. Chou, P. Lodahl, and H. J. Kimble,
      Phys. Rev. A.  {\bf 67}, 033802 (2003).

    \bibitem{Takei05e}
       N. Takei, H. Yonezawa, T. Aoki, and A. Furusawa,
     Phys. Rev. Lett. {\bf 94}, 220502 (2005).

    \bibitem{Yonezawa_QTS(2007)}
       H. Yonezawa, S. L. Braunstein, and A. Furusawa,
      Phys. Rev. Lett. {\bf 99}, 110503 (2007).

    \bibitem{Yukawa08}
     M. Yukawa, H. Benichi, and A. Furusawa,
     Phys. Rev. A.  {\bf 77}, 022314 (2008).

    \bibitem{Braunstein00}
       S. L. Braunstein, C. A. Fuchs, and H. J. Kimble,
       J. Mod. Opt. {\bf 47}, 267 (2000).

    \bibitem{Hammerer05}
      K. Hammerer, M. M. Wolf, E. S. Polzik, and J. I. Cirac,
      Phys. Rev. Lett. {\bf 94}, 150503 (2005).

    \bibitem{Yonezawa_SQT(2007)}
      H. Yonezawa, A. Furusawa, and P. van Loock,
      Phys. Rev. A. {\bf 76}, 032305 (2007).

    \bibitem{Leohnardt97}
      U. Leonhardt,
     \textit{Measuring the Quantum State of Light}, 
     (Cambridge University Press, Cambridge, U.K., 1997).

    \bibitem{Breitenbach97}
    G. Breitenbach and S. Schiller, 
     J. Mod. Opt. {\bf 44}, 2207 (1997).

   \bibitem{Bartlet02}
      S. D. Bartlett, B. C. Sanders, S. L. Braunstein, and K. Nemoto
      Phys. Rev. Lett. {\bf 88}, 097904 (2002).

   \bibitem{Holland90}
      M.J. Holland, M. J. Collett, and D. F. Walls,
      Phys. Rev. A. {\bf 42}, 2995 (1990).

   \bibitem{BachorAGuide}
     Hans-A. Bachor and T. C. Ralph,
    \textit{A Guide to Experiments in Quantum Optics 2nd ed.} 
     (WILEY-VCH, 2004).

    \bibitem{Duan00}
     L.M. Duan, G. Giedke, J. I. Cirac, and P. Zoller,
     Phys. Rev. Lett. {\bf 84}, 2722 (2000).

    \bibitem{Simon00}
     R. Simon,
     Phys. Rev. Lett. {\bf 84}, 2726 (2000).

    \bibitem{vanLoock07}
     P. van Loock, C. Weedbrook, and M. Gu,
     Phys. Rev. A. {\bf 76}, 032321 (2007).

    \bibitem{vanLoock03}
     P. van Loock and A. Furusawa,
     Phys. Rev. A. {\bf 67}, 052315 (2003).

\end{thebibliography}
\end{document}